\documentclass[longauth]{aa} 

\usepackage{txfonts}
\usepackage{graphicx}
\usepackage{amssymb}
\usepackage{natbib}
\usepackage{natbib,amssymb,graphicx,graphics}
\bibpunct{(}{)}{;}{a}{}{,} 
\usepackage{aas_macros}  

\usepackage{multirow}

\begin{document} 

\title{Discovery of a warm, dusty giant planet around HIP\,65426 
\thanks{Based on observations collected at La Silla and Paranal Observatory, ESO (Chile)
Program ID: 097.C-0865 and 098.C-0209 (SPHERE).}
}


\author{G. Chauvin\inst{1,2}, S. Desidera\inst{3}, A.-M. Lagrange\inst{1}, A. Vigan\inst{4}, R. Gratton\inst{3}, M. Langlois\inst{4,5}, M. Bonnefoy\inst{1}, J.-L. Beuzit\inst{1}, M. Feldt\inst{6}, D. Mouillet\inst{1}
% ...
}


\authorrunning{SPHERE consortium}

%
%
        
   \institute{
$^{1}$ Univ. Grenoble Alpes, CNRS, IPAG, F-38000 Grenoble, France. \\
$^{2}$ Unidad Mixta Internacional Franco-Chilena de Astronom\'{i}a, CNRS/INSU UMI 3386 and Departamento de Astronom\'{i}a, Universidad de Chile, Casilla 36-D, Santiago, Chile\\
$^{3}$ INAF - Osservatorio Astronomico di Padova, Vicolo dell’ Osservatorio 5, 35122, Padova, Italy\\
$^{4}$ Aix Marseille Universit\'e, CNRS, LAM (Laboratoire d'Astrophysique de Marseille) UMR 7326, 13388 Marseille, France\\
$^{5}$ CRAL, UMR 5574, CNRS, Universit de Lyon, Ecole Normale Suprieure de Lyon, 46 Alle d'Italie, F-69364 Lyon Cedex 07, France\\
$^{6}$ Max Planck Institute for Astronomy, K\"onigstuhl 17, D-69117 Heidelberg, Germany\\
$^{7}$  Institute for Astronomy, ETH Zurich, Wolfgang-Pauli-Strasse 27, 8093 Zurich, Switzerland\\
$^{8}$ The University of Michigan, Ann Arbor, MI 48109, USA\\
$^{9}$  Geneva Observatory, University of Geneva, Chemin des Mailettes 51, 1290 Versoix, Switzerland\\
$^{10}$ SUPA, Institute for Astronomy, The University of Edinburgh, Royal Observatory, Blackford Hill, Edinburgh, EH9 3HJ, UK\\
$^{11}$ LESIA, Observatoire de Paris, PSL Research University, CNRS, Sorbonne Universités, UPMC Univ. Paris 06, Univ. Paris Diderot, Sorbonne Paris Cité, 5 place Jules Janssen, 92195 Meudon, France\\
$^{12}$ Instituto de F\'isica y Astronom\'ia, Facultad de Ciencias, Universidad de Valpara\'iso, Av. Gran Breta\~na 1111, Valpara\'iso, Chile\\
$^{13}$  Universite Cote d’Azur, OCA, CNRS, Lagrange, France\\
$^{14}$  Department of Astronomy, Stockholm University, AlbaNova University Center, 106 91 Stockholm, Sweden\\
$^{15}$ Physikalisches Institut, University of Bern, Sidlerstrasse 5, 3012 Bern, Switzerland\\
$^{16}$ INAF-Catania Astrophysical Observatory, via S.Sofia, 78 I-95123 Catania, Italy\\
$^{17}$ Núcleo de Astronomía, Facultad de Ingeniería, Universidad Diego Portales, Av. Ejercito 441, Santiago, Chile\\
$^{18}$ Anton Pannekoek Institute for Astronomy, Science Park 904, NL-1098 XH Amsterdam, The Netherlands\\
$^{19}$ INAF - Osservatorio Astrofisico di Arcetri, Largo E. Fermi 5, I-50125 Firenze, Italy\\
$^{20}$ INAF - Osservatorio Astronomico di Capodimonte, Salita Moiariello 16, 80131 Napoli, Italy\\ 
$^{21}$ ONERA (Office National d’Etudes et de Recherches Aérospatiales), B.P.72, F-92322 Chatillon, France\\
$^{22}$ European Southern Observatory (ESO), Karl-Schwarzschild-Str. 2, 85748 Garching, Germany\\
$^{23}$ NOVA Optical Infrared Instrumentation Group, Oude Hoogeveensedijk 4, 7991 PD Dwingeloo, The Netherlands\\
$^{24}$ INAF-Osservatorio Astronomico di Roma, Via di Frascati 33, I-00040 Monte Porzio Catone, Italy\\
$^{25}$ Department of Astrophysics, Denys Wilkinson Building, Keble Road, Oxford, OX1 3RH, UK\\
$^{26}$ European Southern Observatory (ESO), Alonso de Córdova 3107, Vitacura, Casilla 19001, Santiago, Chile\\
$^{27}$ Institute of Astronomy, University of Cambridge, Madingley Road, Cambridge CB3 0HA, UK\\
$^{28}$ School of Earth \& Space Exploration, Arizona State University, Tempe AZ 85287, USA\\
$^{29}$ Ural Federal University, Yekaterinburg 620002, Russia\\
$^{30}$ CNRS, CFHT, 65-1238 Mamalahoa Hwy, Kamuela HI 96743, USA
$^{31}$ Exoplanets and Stellar Astrophysics Laboratory, NASA Goddard Space Flight Center, 8800 Greenbelt Rd., Greenbelt, MD 20771, USA\\
             \email{gael.chauvin@univ-grenoble-alpes.fr}
             }

   \date{Received May 10th, 2017}

 
  \abstract
   {}
     {The SHINE program is a high-contrast near-infrared survey
       of 600 young, nearby stars aimed at searching for and
       characterizing new planetary systems using VLT/SPHERE's
       unprecedented high-contrast and high-angular-resolution imaging
       capabilities. It is also intended to place statistical constraints
       on the rate, mass and orbital distributions of the giant planet
       population at large orbits as a function of the stellar host
       mass and age to test planet-formation theories.}
      {We used the IRDIS dual-band imager and the IFS integral field
        spectrograph of SPHERE to acquire high-contrast coronagraphic differential near-infrared
        images and spectra of the young A2 star HIP\,65426. It is a member
        of the $\sim17\,$Myr old Lower Centaurus-Crux association.}
    {At a separation of 830\,mas (92 au projected) from the star, we
      detect a faint red companion. Multi-epoch observations
      confirm that it shares common proper motion with
      HIP\,65426. Spectro-photometric measurements extracted with IFS
      and IRDIS between 0.95 and 2.2\,$\mu$m indicate a warm, dusty
      atmosphere characteristic of young low-surface-gravity L5-L7
      dwarfs. Hot-start evolutionary models predict a luminosity consistent with
      a $6-12$\,$M_{\rm{Jup}}$,
      $T_{\rm{eff}}=1300-1600$\,K and
      $R=1.5\pm0.1\,R_{\rm{Jup}}$ giant planet.  Finally, the comparison with Exo-REM and PHOENIX BT-Settl synthetic
      atmosphere models gives consistent effective temperatures but with slightly higher surface gravity solutions
of $log(g)=4.0-5.0$ with smaller radii ($1.0-1.3\,R_{\rm{Jup}}$).}
   {Given its physical and spectral properties, HIP\,65426\,b occupies
     a rather unique placement in terms of age, mass, and spectral-type
     among the currently known imaged planets. It represents a
     particularly interesting case to study the presence of clouds as
     a function of particle size, composition, and location in the
     atmosphere, to search for signatures of non-equilibrium
     chemistry, and finally to test the theory of planet formation and
     evolution.}
   \maketitle
%
\section{Introduction}


More than a decade of direct imaging surveys targeting several
hundred young, nearby stars have revealed that the occurrence of
giant planets at wide orbits ($\ge20-40~au$) is relatively low
\citep[e.g.][]{bowler2016}. Despite the relatively small number of
discoveries compared with other techniques, such as radial velocity and
transit, each new imaged giant planet has provided unique clues on the
formation, evolution and physics of young Jupiters. The latest
generation of planet imagers, SPHERE \citep{beuzit2008}, GPI
\citep{macintosh2014} and SCExAO \citep{jovanovic2016}, now combine
innovative extreme adaptive optics systems with coronagraphic and
differential imaging techniques. They offer unprecedented detection,
astrometric and spectrophotometric capabilities which allow us to
discover and characterize fainter and closer giant planets, such as the
recent discovery of 51\,Eri\,b (2~M$_{\rm{Jup}}$ at 14~au,
T$5$-type, of age 20~Myr; \citealt{macintosh2015}; Samland et
al. 2017). The SHINE (SpHere INfrared survey for Exoplanets) survey is currently
surveying 600 young, nearby stars as part of the SPHERE Guaranteed Time
Observations. In this survey, we observed the close environment of the
young, star HIP\,65426.  The deep coronographic near-infrared observations
revealed the presence of a young, warm, and dusty L5-L7 massive jovian
planet, hereafter HIP\,65426\,b, located at about 92\,au (projected
distance).  We describe below the observing set-up and data reduction,
the physical properties of HIP\,65426\,b, and finally discuss this new
discovery in comparison to other imaged planetary systems and current
planet formation and evolution theories.

\section{Host star properties}
\label{sec:host}

HIP\,65426 is an A2-type ($H =
  6.853\pm0.049$\,mag, \citealt{cutri2003}; $d=111.4\pm3.8$\,pc,
  \citealt{gaia2016}) member of the Lower Centaurus-Crux (hereafter
LCC) association \citep{dezeeuw1999,rizzuto2011}. A detailed summary
of the main stellar properties as found in the literature is given in
Appendix\,\ref{s:isoc}. To refine them, the star was observed with HARPS
\citep{mayor2003} on January 16th, 17th and 18th, 2017\footnote{HARPS
  Program ID 098.C-0739(A)}. We measured the stellar absolute radial
and projected rotational velocities using a custom cross-correlation
function procedure specifically tailored for fast-rotating early-type
stars. Values of $V_{rad}=5.2\pm1.3$\,km.s$^{-1}$ and
$v\sin i=299\pm9$\,km.s$^{-1}$ were found (Appendix\,\ref{s:rot}). A
marginally significant radial velocity difference was found between the three
epochs. This is probably due to stellar pulsations as suggested by the periodicity
($P\sim0.135$\,days) found from the Hipparcos photometric time
series. HIP\,65426 is one of the fastest rotators known with similar
spectral type \citep{zorec2012} and is therefore likely viewed
along mid- to high-inclinations with respect to the rotation axis.  Given its
spectral type, the observed colors of HIP\,65426 suggest a small value
of reddening consistent with estimates reported in \cite{chen2012}. Assuming a
metallicity for the LCC that is close to solar \citep{vianaalmeida2009},
theoretical isochrones predict an age of 14$\pm$4\,Myr for LCC members
in the vicinity of HIP\,65426 (we refer to Appendix\,\ref{s:isoc} for details). SPHERE and
HARPS observations do not show evidence of binarity
(Appendix\,\ref{s:bin}).  Finally, according to \cite{chen2012}, no IR
excess is reported for this star. Our own SED analysis confirms this
finding with only a tentative marginal excess at WISE W4 (Appendix\,\ref{s:excess}).

\section{Observations and data reduction}
\label{sec:obs}

HIP\,65426 was observed on May 30th, 2016 under unstable conditions
(strong wind) with SPHERE. The observations were then repeated on June
26th, 2016.  The data were acquired in IRDIFS pupil-tracking mode with
the 185\,mas diameter apodized-Lyot coronograph
\citep{carbillet2011,guerri2011}, using IRDIS
\citep{dohlen2008} in dual-band imaging mode \citep{vigan2010} with
the $H_2H_3$ filters ($\lambda_{H_2} = 1.593\pm0.055~\mu$m;
$\lambda_{H_3} = 1.667 \pm 0.056~\mu$m), and the IFS integral field
spectrograph \citep{claudi2008} simultaneously in $Y-J$ ($0.95-1.35\,\mu$m,
$R_\lambda=54$) mode. The registration of the star
position behind the coronagraph and the point spread function were
taken at the beginning and the end of the sequence. In the deep
coronagraphic images, four faint candidates were detected within $7\,\!''$ of HIP\,65426.  The companion candidate located at about
$830$~mas and position angle of $150\degr$ (hereafter cc-0) revealed promising
photometric properties and had a peculiar position in the
color-magnitude diagrams used to rank the SHINE candidates. The source
was then re-observed on February 7th, 2017, with the same IRDIFS mode
to test that this close candidate is comoving with HIP\,65426.  On
February 9th, 2017, the IRDIFS-EXT mode was used with IRDIS in the
$K_1K_2$ filters ($\lambda_{K_1} = 2.1025 \pm 0.1020\,\mu$m;
$\lambda_{K_2} = 2.2550 \pm 0.1090\,\mu$m) and IFS in $Y-H$
($0.97-1.66\,\mu$m, $R_\lambda=30$) to further constrain its
physical and spectral properties. The details of the observing
settings and conditions at all epochs are described in
Table~\ref{tab:obslog} of Appendix~\ref{s:obslog}. To calibrate the IRDIS and IFS datasets, an
astrometric field 47\,Tuc was observed. The platescale and True North
correction solution at each epoch are reported in Table~\ref{tab:obslog}. They
are derived from the long-term analysis of the SHINE astrometric
calibration described by \cite{maire2016}.

\begin{figure}[hb]
\hspace{0.2cm}
\includegraphics[width=\columnwidth]{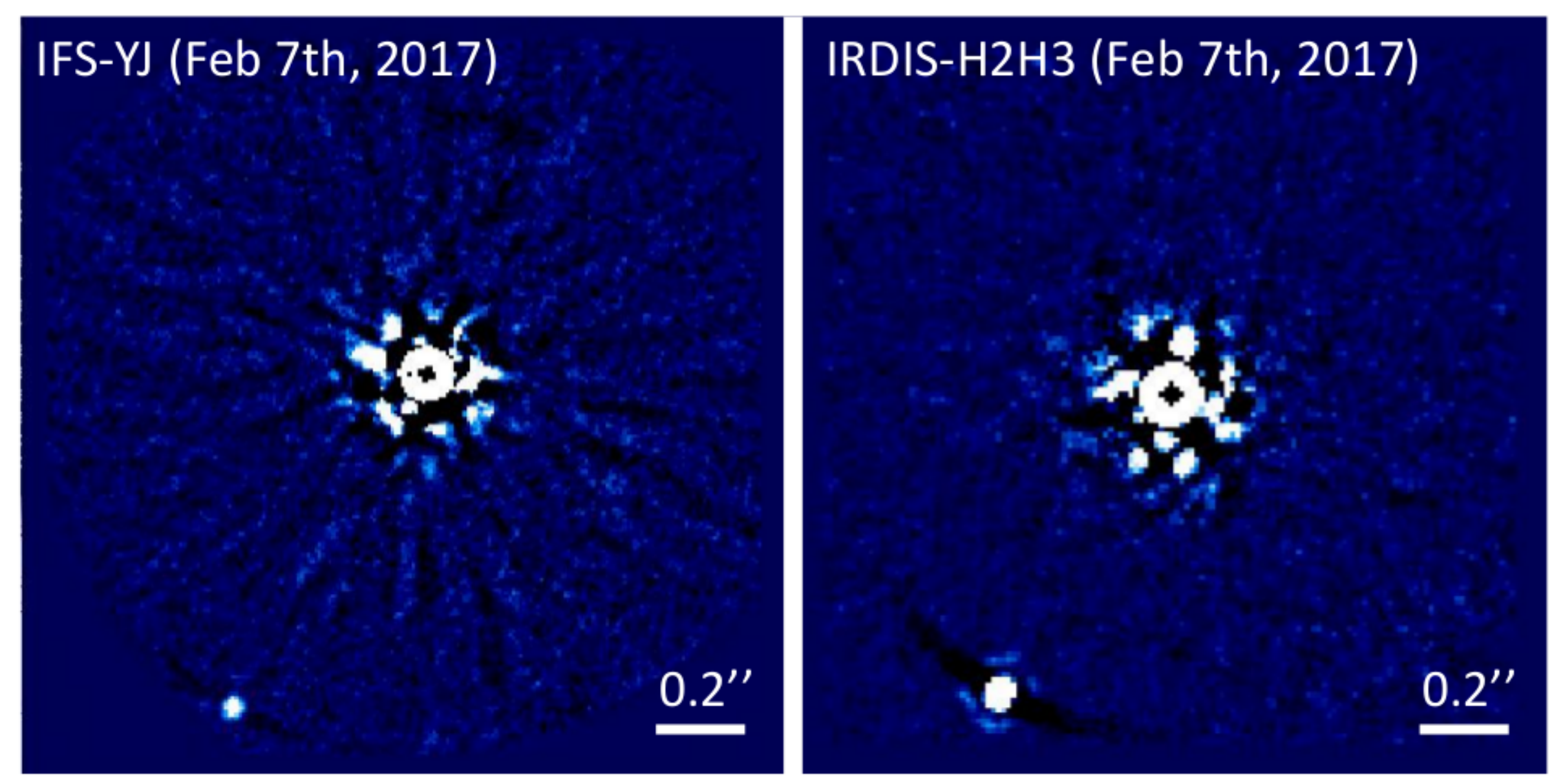}
\begin{centering}
\caption{\textit{Left:} IFS $YJ$-band TLOCI image of
  HIP\,65426\,A and b from February 7th 2017. The planet is well detected at a separation of
  $830\pm3$~mas and position angle of $150.0\pm0.3$~deg from
  HIP\,65426. \textit{Right:} IRDIS $H_2H_3$ combined TLOCI image of
  HIP\,65426\,A and b for the same night. In both images, North is up and East is left.}
\label{fig:image}
\end{centering}
\end{figure}

All IRDIS and IFS datasets were reduced at the SPHERE Data
Center\footnote{http://sphere.osug.fr} (DC) using the SPHERE Data
Reduction and Handling (DRH) automated pipeline
\citep{pavlov2008}. Basic corrections for bad pixels, dark current,
flat field were applied. For IFS, the SPHERE-DC complemented the DRH
pipeline with additional steps that improve the wavelength calibration
and the cross-talk and bad pixel correction (Mesa et al. 2015). The
products were then used as input to the SHINE \textit{Specal} pipeline
which applies anamorphism correction and flux normalization, followed
by different angular and spectral differential imaging
algorithms (Galicher et al. 2017, in prep). For the February 7th, 2017
and February 9th, 2017 datasets, we took advantage of the waffle-spot
registration to apply a frame-to-frame recentering. The TLOCI
\citep{marois2014} and PCA \citep{soummer2012,amara2012} algorithms
were specifically used on angular differential imaging data (i.e.,
without applying any combined spectral differential processing) given
the relatively high S/N ($10-30$ in the individual IFS channels,
$\ge50$ with IRDIS) detection of cc-0. Its position and
spectrophotometry were extracted using injected fake planets and planetary
signature templates to take into account any biases related to the
data processing.  Both algorithms gave consistent results.  The
resulting extracted TLOCI images from IFS and IRDIS for the February
7th, 2017 epoch are shown in Fig.\,\ref{fig:image}.

%
\section{Results}
\label{sec:results}

\subsection{Companionship confirmation}
\label{subsec:astro}

\begin{figure}[t]
\hspace{0.2cm}
\includegraphics[width=8cm]{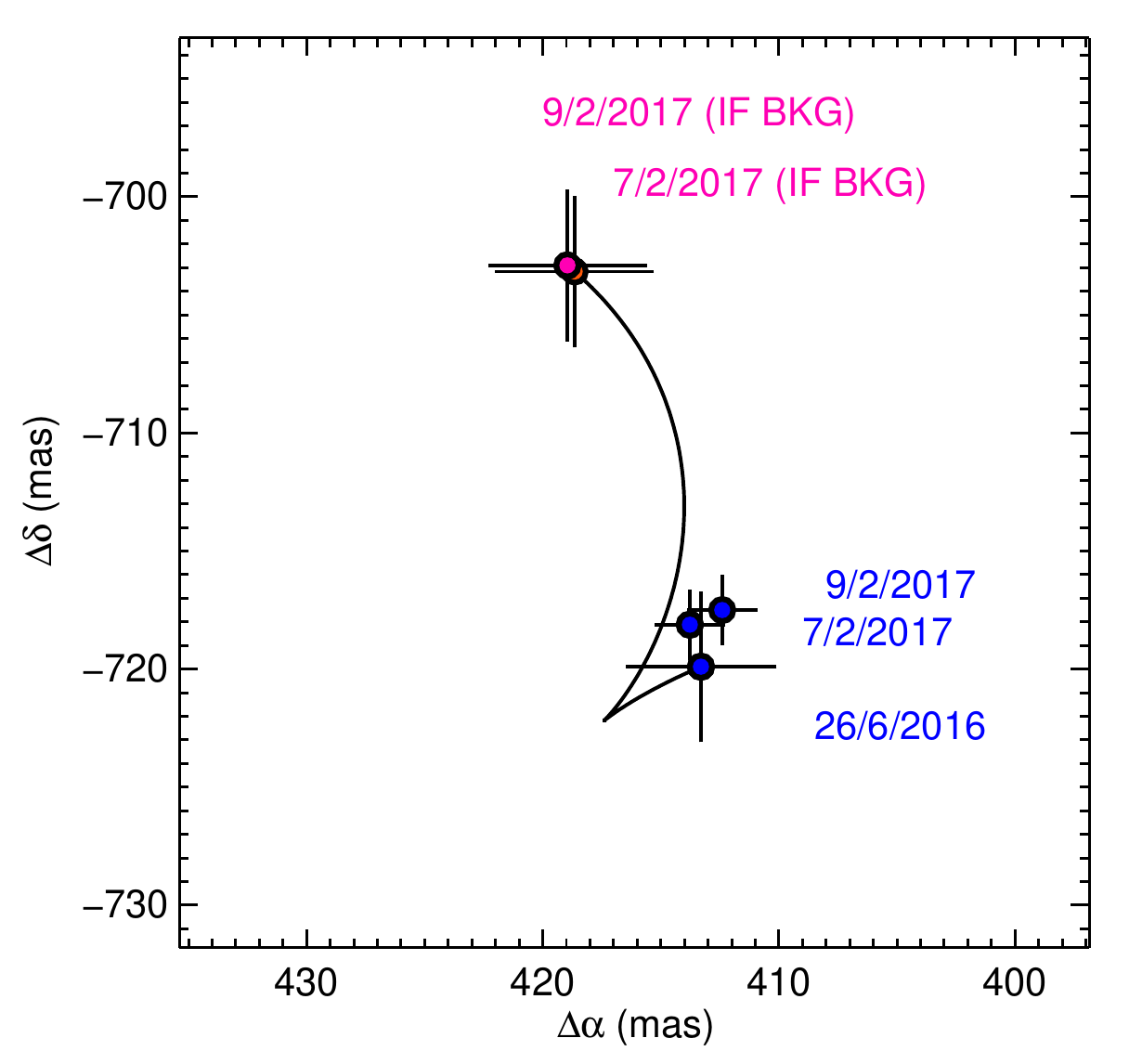}
\begin{centering}
\caption{
IRDIS multi-epoch measurements of the position of HIP\,65426\,b
relative to HIP\,65426 in \textit{blue} from June 26th, 2016 and
February 7th, 2017 in $H_2$ and  February 9th,
2017 in $K_1$. Predictions of the relative position of a
stationary background contaminant for the same observing epochs are shown in \textit{pink} and in \textit{black} for the continuous evolutive predictions in time. }
\label{fig:motion}
\end{centering}
\end{figure}

To test the physical association of cc-0, multi-epoch measurements
showing a shared motion (if non-negligible) with the stellar host is a
first robust diagnostic before resolving orbital motion. We therefore
used the IRDIS observations of June 26th, 2016, and February 7th and
9th, 2017. We did not consider the first observation of the May 30th 2016
epoch which was acquired under unstable conditions.  The astrometric
uncertainties are derived at each epoch by quadratically summing
errors from the stellar position calibrated with the waffle-spots, the
candidate extracted position with Specal and the uncertainties related
to the initial pupil-offset rotation, the platescale and True North correction,
the anamorphism and the IRDIS dithering. The photometric
  results obtained with both IRDIS and IFS are given in
  Table\,\ref{tab:astro1} of Appendix\,\ref{s:astrophot} considering the SPHERE filter
  transmissions\footnote{https://www.eso.org/sci/facilities/paranal/instruments/sphere/inst/filters.html}.
The astrometric results obtained with IRDIS are reported in
Table\,\ref{tab:astro1} and in Fig.\,\ref{fig:motion}. Consistent
astrometric results were found with IFS. They both unambiguously
confirm that cc-0 is not a stationary background contaminant
but is comoving with HIP\,65426 and is therefore likely a planet
(hereafter HIP\,65426\,b). One additional close candidate
  (cc-1) located at 2495\,mas provides a robust control of our
  astrometric analysis and is confirmed as a stationary background
  contaminant. The nature of the two other candidates (cc-2 and cc-3)
  at larger separations remains to be clarified (considering the
  astrometric error bars), but their position in the color-magnitude
  diagram indicates that they are very likely background objects (see Fig.\,\ref{fig:cmd} of \,Appendix\,\ref{s:astrophot}). The
  relative astrometry and photometry for these additional candidates
  are reported in Table\,\ref{tab:astro2} of Appendix\,\ref{s:astrophot} for June 26th, 2016. No
significant orbital motion for HIP\,65426\,b is measured in the
present data. This is consistent with the expected long orbital period
($P\sim600$\,yr for a circular orbit with semi-major axis equal to the
projected separation). Finally, the probability of having at
  least one background contaminant of similar brightness or brighter
  within the separation of HIP\,65426\,b and with proper motion less
  than 5\,$\sigma$ deviant from the HIP\,65426 proper motion is less
  than 1\% given the galactic coordinates of HIP\,65426 and the
  predicted space and velocity distribution of field stars from the
  \textit{Galaxia} galactic population model of \cite{sharma2011}.


\subsection{Spectral typing analysis}
\label{subsec:spectra}

The TLOCI extracted spectro-photometric measurements of HIP\,65426\,b
between 0.95 and 2.26\,$\mu$m (converted to physical fluxes using the
VOSA\footnote{http://svo2.cab.inta-csic.es/theory/vosa/} tool) are
reported in Fig.\,\ref{fig:spectra}. We compared them to a large
variety of reference low-resolution spectra of late-M and L dwarfs
compiled from the literature
\citep{burgasser2014,best2015,mace2013,allers2013} as well as spectra
of young imaged exoplanets and brown dwarfs close to the L/T
transition
\citep{patience2010,zurlo2016,derosa2014,artigau2015,gauza2015}. We
considered the $G$ goodness-of-fit indicator defined in Cushing et
al. (2008) which accounts for the filter and spectral channel widths
to compare each of the template spectra to the spectrophotometric
datapoints of HIP\,65426\,b. The best empirical fits are obtained for
the young L5 and L7 dwarfs 2MASS\,J035523.37+113343.7 and
PSO\,J057.2893+15.2433 recently identified as candidate members of the
young, moving groups AB Doradus ($50-150$\,Myr) and $\beta$ Pictoris
($20-30$\,Myr), respectively, \citep{faherty2013,liu2013,best2015} and
the dusty L6 dwarf 2MASS\,J21481628+4003593
\citep{looper2008}. Fig.\,\ref{fig:spectra} shows how well they
reproduce the near-infrared slope of the spectrum of HIP\,65426\,b
between 0.95 and 2.26\,$\mu$m as well as the water absorption at
$1.33$\,$\mu$m. This comparison confirms a low-surface-gravity
atmosphere of spectral type L$6\pm1$ for HIP\,65426\,b consistent with
a young massive planet at the age of the LCC.

\subsection{Physical properties}
\label{subsec:physics}

\begin{figure}[t]
\includegraphics[width=\columnwidth]{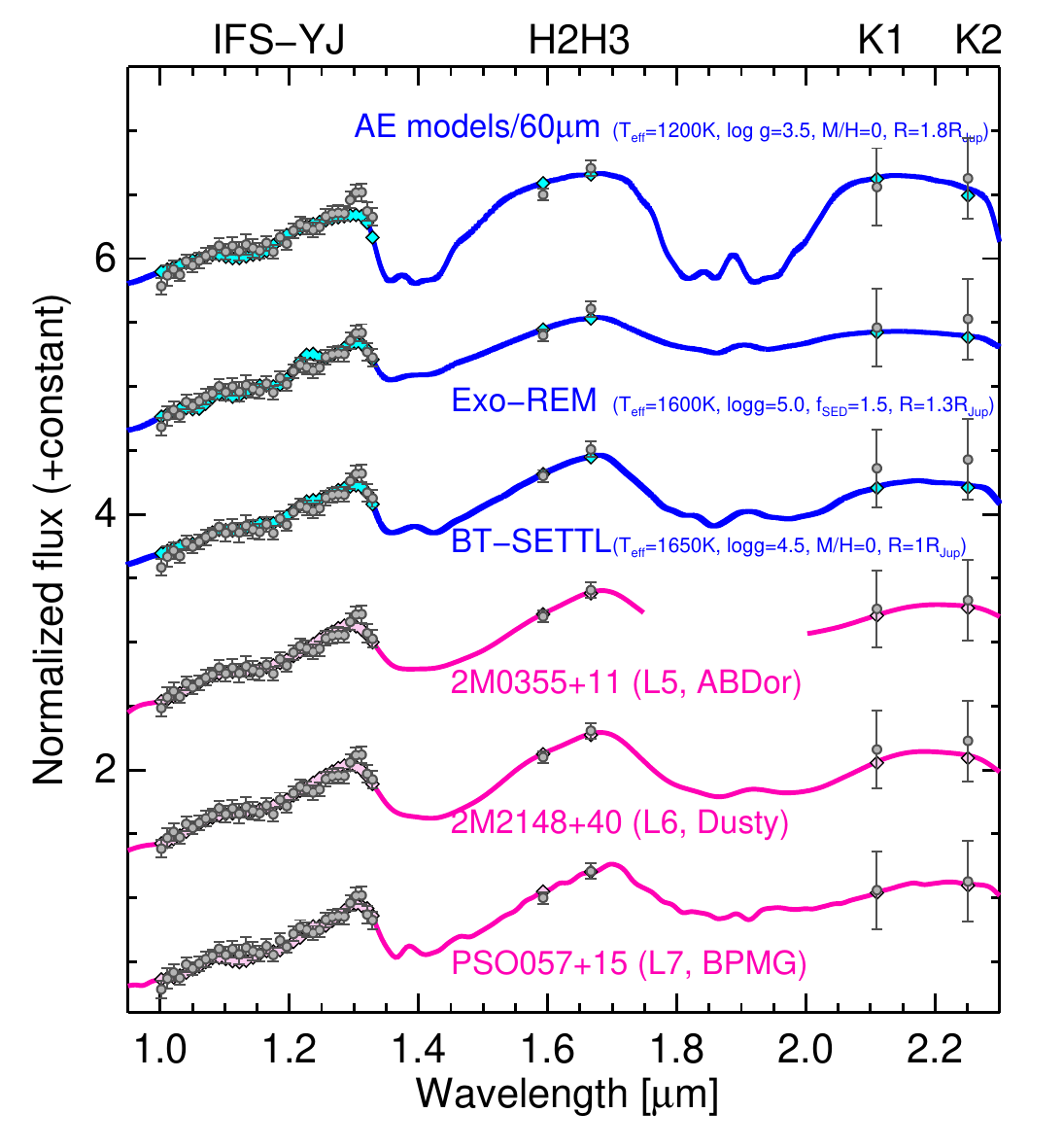}

\caption{Near-infrared spectrum of HIP\,65426\,b extracted with TLOCI
  compared with (i) the best-fit empirical spectra in \textit{pink},
  and (ii) the best-fit model atmosphere from the Exo-REM, PHOENIX
  BT-Settl-2014 and thick AE cloud atmospheric models in
  \textit{blue}.  }
\label{fig:spectra}
\end{figure}

Using a bolometric correction derived from the ones found by
  \citep{filippazzo2015} for the dusty L5 to L7.5 dwarfs
  2MASS\,J21481628+4003593, 2MASS\,J035523.37+113343.7,
  PSO\,J318.5338–22.8603, and WISEP\,J004701.06+680352.1, we derive a
  bolometric luminosity of $-4.06\pm0.10\,$dex for
  HIP\,65426\,b. Considering an age of $14\pm4$\,Myr and a distance of
  $111.4\pm3.8$\,pc, it converts into the following predicted masses,
  effective temperatures and radii by the Lyon's group hot-start models:
  $M=7_{-1}^{+2}\,M_{\rm{Jup}}$, $T_{\rm{eff}}=1500_{-200}^{+100}\,$K
  and $R=1.5\pm0.1\,R_{\rm{Jup}}$ for the COND03 models
  \citep{baraffe2003} and $M=10\pm2\,M_{\rm{Jup}}$,
  $T_{\rm{eff}}=1500_{-200}^{+100}\,$K and
  $R=1.5\pm0.1\,R_{\rm{Jup}}$ for the DUSTY models
  \citep{chabrier2000}. Consistent results are found with the models
of \citet{mordasini2013} for the hot-start solutions and higher masses
for the warm-start solutions ($M=12\,M_{\rm{Jup}}$,
$T_{\rm{eff}}=1260\,$K and $R=1.3\,R_{\rm{Jup}}$). To further
explore the physical properties of HIP\,65426\,b, we compared our data
to the synthetic grids of three atmospheric models previously used in
the characterization of the planets around HR\,8799
\citep{bonnefoy2016}. They are the Exo-REM models \citep{baudino2015},
the 2014 version of the PHOENIX BT-Settl atmospheric models described
in \cite{allard2014} and \cite{baraffe2015} and the thick AE cloud
parametric models of Madhusudhan et al. (2011). The best fits for each
grid are reported in Fig.\,\ref{fig:spectra}. The Exo-REM and
PHOENIX BT-Settl models favor an effective temperature of
$T_{\rm{eff}}=1600_{-200}^{+100}$\,K, slightly higher than the one
derived by the semi-empirical scale of \citep{filippazzo2015} for
young L$6\pm1$ ($T_{\rm{eff}}=1200-1400$\,K), but consistent with the evolutionary
model predictions. They however favor high-surface-gravity solutions
of $log(g)=4.0-5.0$ with smaller radii ($1.0-1.3\,R_{\rm{Jup}}$). On
the contrary, the thick AE cloud parametric models that predict
solutions with $T_{\rm{eff}}=1200\pm100$\,K and $log(g)=3.5$ reproduce
the spectral morphology but predict luminosities which are too low,
thus leading to radii above the evolutionary model predictions
($\sim1.8\,R_{\rm{Jup}}$). 

The inferred chemical and physical properties of HIP\,65426\,b place
this new planet in a very interesting sequence of young brown dwarfs
and exoplanets discovered in the $5-20$\,Myr-old Scorpius-Centaurus
association (hereafter Sco-Cen). It is lighter and cooler than the
late-M brown dwarf companions discovered by
\cite{aller2013} and \cite{hinkley2015} and the massive planetary mass companions
GSC\,06214-00210\,b ($14-17\,M_{\rm{Jup}}$ at 320~au; M9 at
  5~Myr; \citealt{lachapelle2015,ireland2011}), UScoCTIO\,108\,b
  ($6-16~M_{\rm{Jup}}$ at 670~au; M9.5 at 5~Myr;
  \citealt{bonnefoy2014,bejar2008}), HD\,106906\,b ($11~M_{\rm{Jup}}$
  at 650~au; L$2.5$ at 13~Myr; \citealt{bailey2014}) and
  1RXS\,J160929.1-210524\,b ($7-12~M_{\rm{Jup}}$ at 330~au; L$2-4$ at
  5~Myr; \citealt{lachapelle2015,manjavacas2014,lafreniere2008}). On
the other hand, HIP\,65426\,b is probably more massive and hotter than
the planet HD\,95086\,b ($4-5~M_{\rm{Jup}}$ at 56~au, L$8$-type,
17~Myr; \citealt{rameau2013b}). This spectral and physical sequence is
particularly interesting to study the main phase of transitions
occurring in the atmosphere of brown dwarfs and exoplanets and
influencing their spectra and luminosity, such as the formation of
clouds and their properties as a function of particle size,
composition, and location in the atmosphere or the role of
non-equilibrium chemistry processes. Further characterization in the
thermal-infrared domain with \textit{JWST} or ground-based instrument
like NaCo will allow us to explore in more detail the young planetary
atmosphere of HIP\,65426\,b. Dedicated photometric variability
monitoring would also be opportune as HIP\,65426\,b shares a similar
spectral type and young age with the two highest-amplitude (7-10\%)
variable L-type dwarfs known (PSO\,J318.5-22, L7 member of
  $\beta$ Pic, \citealt{biller2015,liu2013}) and WISE\,0047, L6.5
  member of AB Dor, \citealt{lew2016,gizis2012}) and, as radial
velocity measurements suggest, we may be observing the system
close to edge-on.

%
\section{Discussion}
\label{sec:discu}


\begin{figure}[t]
\includegraphics[width=\columnwidth]{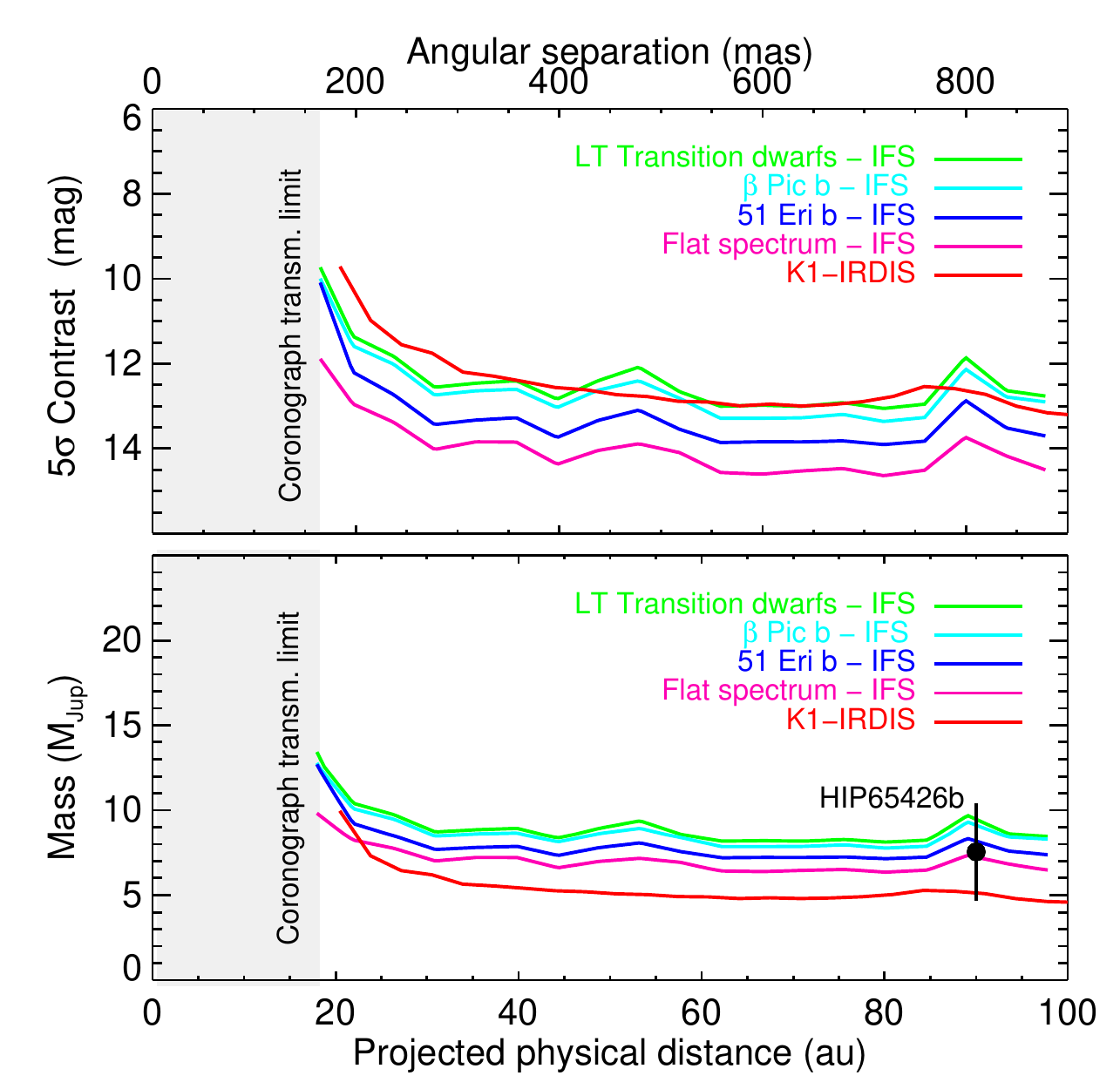}\\

\caption{\textit{Top}: SPHERE IFS and IRDIS $5\sigma$ detection limits
  as a function of the angular separation taken from February 7th and
  9th, 2017. \textit{Bottom}: SPHERE IFS and IRDIS $5\sigma$ detection
  limits converted in terms of masses using DUSTY model predictions as
  a function of the projected physical separation. For IFS, different
  spectral energy distributions were considered for the injected
  planets to explore the impact of the flux loss cancellation and
  different planet properties in the final detection limits. Contrast
  curves were cut at $0.15\,\!''$ because of the low transmission of
  the coronagraph. The location and the predicted mass by the DUSTY models of
  HIP\,65426\,b are reported.}

\label{fig:detlim}
\end{figure}

HIP\,65426\,b is today the first planet\footnote{With a mass ratio to its host of $q=0.004$, we consider
 HIP\,65426\,b as a planet
as suggested by the local minimum observed for the mass-ratio distribution
 of low-mass companions orbiting Sun-like stars \citep{sahlmann2011,reggiani2016}.} discovered with
  SPHERE. The planet is orbiting at a relatively large projected
physical distance of about 92\,au from the intermediate-mass primary
HIP\,65426. Contrary to most of the young, intermediate-mass stars
hosting an imaged planet, no evidence of a debris disk, a remnant planetesimal belt, has been found
for HIP\,65426. The analysis of the optical to mid-infrared photometry
shows that, if the star is still hosting a debris disk, it would be
located at distances larger than 100\,au (i.e. farther out than the
planet location) and with an upper limit to the micron-sized dust mass of $3.2
\times 10^{-4}$\,$M_{\oplus}$ (see Appendix \ref{s:excess}). No signs of
multiplicity have been observed so far for HIP\,65426, which could
have explained a rapid dispersal of the primordial protoplanetary
disk, but this should still be investigated.  Another intriguing
aspect of the system is that HIP\,65426 is an extremely fast rotator
as evidenced by our HARPS observations. No similar cases are known
among the Sco-Cen and young, nearby intermediate-mass association
members, nor among the intermediate-mass primaries hosting young
imaged giant planets (see Appendix \ref{s:rot}).  Although fast stellar
rotation is consistent with the picture of a rapid disk dispersal
disabling disk-braking, planetary formation must have time to occur to
explain the formation of HIP\,65426\,b.  The planet location would not
favor a formation by core accretion unless HIP\,65426\,b formed
significantly closer to the star followed by a planet-planet
scattering event. An increase of angular momentum by engulfing an
inner massive scatterer could explain the fast rotation of HIP\,65426,
but this remains to be tested by dedicated simulations. From
our observations, we cannot exclude the presence of unseen inner
massive planets in that system that could have scattered out
HIP\,65426\,b. However, our current detection limits set relatively
good constraints on their possible masses ($\le5$\,$M_{\rm{Jup}}$
beyond 20\,au), as shown in Fig.\,\ref{fig:detlim}. As a consequence
of a scattering event, the orbit of HIP\,65426\,b would also be rather
eccentric which could be probed with further astrometric
monitoring. If formed in-situ at its current location, formation by
disk instability would be a better alternative, which would be
consistent with the metallicity of the host star not enhanced
  with respect to the solar value. Finally, the formation of an
extreme mass-ratio binary by gravo-turbulent fragmentation
\citep{hennebelle2011} cannot be excluded.




\bibliographystyle{aa}

\begin{acknowledgements}

We acknowledge financial support from the Programme National de
Planétologie (PNP) and the Programme National de Physique Stellaire
(PNPS) of CNRS-INSU. This work has also been supported by a grant from
the French Labex OSUG@2020 (Investissements d’avenir – ANR10 LABX56).
The project is supported by CNRS, by the Agence Nationale de la
Recherche (ANR-14-CE33-0018). This work has made use of the the SPHERE
Data Centre, jointly operated by OSUG/IPAG (Grenoble),
PYTHEAS/LAM/CESAM (Marseille), OCA/Lagrange (Nice) and Observtoire de
Paris/LESIA (Paris). We thank P. Delorme and E. Lagadec (SPHERE Data
Centre) for their efficient help during the data reduction
process. SPHERE is an instrument designed and built by a consortium
consisting of IPAG (Grenoble, France), MPIA (Heidelberg, Germany), LAM
(Marseille, France), LESIA (Paris, France), Laboratoire Lagrange
(Nice, France), INAF–Osservatorio di Padova (Italy), Observatoire de
Genève (Switzerland), ETH Zurich (Switzerland), NOVA (Netherlands),
ONERA (France) and ASTRON (Netherlands) in collaboration with
ESO. SPHERE was funded by ESO, with additional contributions from CNRS
(France), MPIA (Germany), INAF (Italy), FINES (Switzerland) and NOVA
(Netherlands).  SPHERE also received funding from the European
Commission Sixth and Seventh Framework Programmes as part of the
Optical Infrared Coordination Network for Astronomy (OPTICON) under
grant number RII3-Ct-2004-001566 for FP6 (2004–2008), grant number
226604 for FP7 (2009–2012) and grant number 312430 for FP7
(2013–2016). M. B. thanks A. Best, K. Allers, G. Mace, E. Artigau,
B. Gauza, R. D. Rosa, M.-E. Naud, F.-R. Lachapelle, J. Patience,
J. Gizis, A. Burgasser, M. Liu, A. Schneider, K. Aller, B. Bowler,
S. Hinkley, and K. Kellogg for providing their spectra of young, brown
dwarf companions.  This publication makes use of VOSA, developed under
the Spanish Virtual Observatory project supported from the Spanish
MICINN through grant AyA2011-24052. This research has benefitted from
the SpeX Prism Spectral Libraries, maintained by A. Burgasser at
http://pono.ucsd.edu/$\sim$adam/browndwarfs/spexprism.
This research has made use of the Washington Double Star Catalog 
maintained at the U.S. Naval Observatory. This work has made use of data from the European Space Agency (ESA)
mission {\it Gaia} (\url{https://www.cosmos.esa.int/gaia}), processed by
the {\it Gaia} Data Processing and Analysis Consortium (DPAC,
\url{https://www.cosmos.esa.int/web/gaia/dpac/consortium}). 
Part of this work has been carried out within the frame of the National Centre for Competence in 
Research PlanetS supported by the Swiss National Science Foundation (SNSF). MRM, HMS, and SD are pleased 
to acknowledge this financial support of the SNSF.

\end{acknowledgements}

%
%

\bibliography{hip65426}

\appendix  
\section{Isochronal ages of HIP\,65426 and neighboring stars}
\label{s:isoc}

\begin{center}
\begin{figure}
\includegraphics[width=9cm]{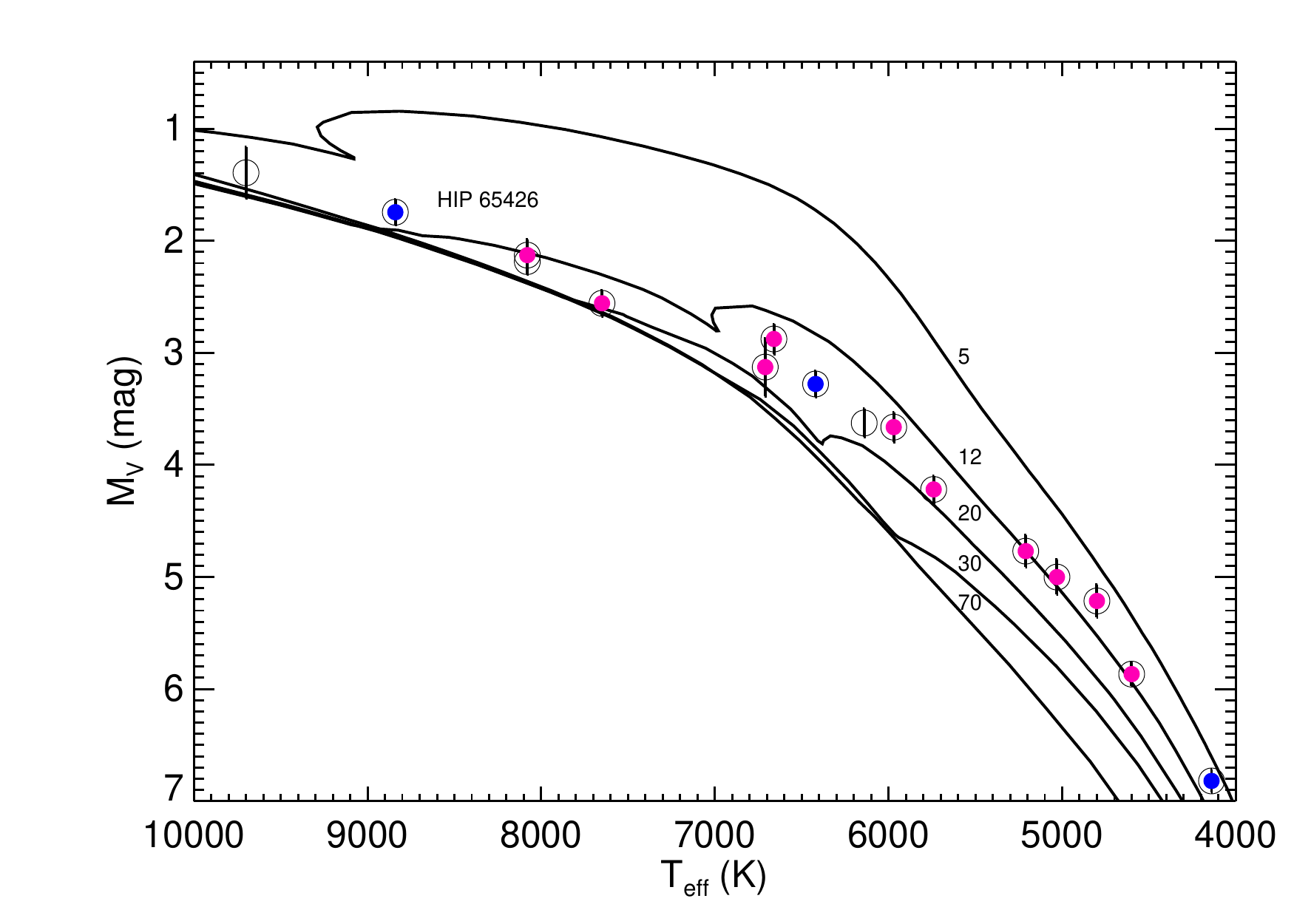}

\caption{Absolute magnitude in $V$-band versus effective temperature
  diagram for HIP 65426 and stars within 5\,\degr\, with similar
  distance and kinematics. \textit{Blue circles}: HIP\,65426 and two
  stars with space velocity difference smaller than 3 km.s$^{-1}$.
  \textit{Red circles}: Stars with space velocity difference with
  respect to HIP 65426 between 3 and 10\,km.s$^{-1}$. \textit{Empty
    circles:} stars with space velocity difference larger than 10
  km.s$^{-1}$.  The 5, 12, 20, 30, 70 Myr solar-metallicity isochrones
  by Bressan et al. (2012) are overplotted and labeled
  individually.} \label{f:isoc}

\end{figure}
\end{center}

Adopting the values from Table~\ref{ref:star_full}, the placement in
absolute magnitude in $V$-band versus effective temperature diagram
and comparison with theoretical models by \cite{bressan2012} yields an
age of 9-10 Myr when adopting the Gaia DR1 parallax (Fig.~\ref{f:isoc}). This is slightly younger than the commonly adopted
age for LCC (17 Myr). However, recent results highlight the existence
of significant age differences at various locations within the Sco-Cen
sub-groups \citep{fang2017,pecaut2016}.

\begin{center}
\begin{table}[h]
\caption{Stellar parameters of HIP 65426.}\label{ref:star_full}
\centering
\begin{tabular}{lcl}
\hline\hline\noalign{\smallskip}
Parameter      & Value  & References \\
\noalign{\smallskip}\hline\noalign{\smallskip} 
$V$ (mag)                   &  7.01       & \textit{Hipparcos}$^a$ \\
$B-V$ (mag)               &  0.093      & \textit{Hipparcos}$^a$ \\
$V-I$ (mag)               &  0.11       & \textit{Hipparcos}$^a$ \\
$J$ (mag)                   &  6.826$\pm$0.019  & \textit{2MASS}$^b$ \\
$H$ (mag)                   &  6.853$\pm$0.049  & \textit{2MASS}$^b$ \\
$K$ (mag)                   &  6.771$\pm$0.018  & \textit{2MASS}$^b$ \\
Parallax (mas)            &    8.98$\pm$0.30  & \textit{Gaia DR1}$^c$ \\
$\mu_{\alpha}$ (mas.yr$^{-1}$)  & -33.923$\pm$0.030  & \textit{Gaia DR1}$^c$ \\
$\mu_{\delta}$ (mas.yr$^{-1}$)  & -18.955$\pm$0.031  & \textit{Gaia DR1}$^c$ \\
RV   (km.s$^{-1}$)            &  -5.2$\pm$1.3  & this paper \\
SpT                            & A2V & \\
$T_{\rm eff}$ (K)      &  8840$\pm$200 & SpT+Pecaut calib. \\
E(B-V)               &   0.01$\pm$0.01 & this paper \\
$v \sin i $  (km.s$^{-1}$)          &   299$\pm$9 & this paper \\
Age (Myr)            &   $14^{+4}_{-4}$ & this paper \\
$M_{star} (M_{\odot})$   &  1.96$\pm$0.04 & this paper \\  
$R_{star} (R_{\odot})$   &  1.77$\pm$0.05 & this paper \\
\noalign{\smallskip}
\hline\noalign{\smallskip}
\end{tabular}
\begin{list}{}{}
\item \scriptsize{References: (a) \citealt{perryman1997}, (b) \citealt{cutri2003}, (c) \citealt{gaia2016}.}
\end{list}
\end{table}
\end{center}

To further refine the age estimate, we considered additional stars
physically close to HIP\,65426 and with similar kinematic parameters.
We selected from Gaia DR1 stars within 5\,\degr\, of HIP\,65426,
with parallax within 1\,mas, and proper motion within
6\,mas.yr$^{-1}$.  This search returned a total of 15 objects. Eight of
these were previously known as LCC members.  The seven stars without
RV determination from the literature were observed with FEROS
spectrograph at 2.2\,m telescope at La Silla as part of MPIA observing
time\footnote{Program ID 098.A-9007(A)}. Full results of these
observations will be presented in a future publication.  All the
selected stars are probable members of the Sco-Cen group, as resulting
from RV, signatures of fast rotation and activity or lithium from data
available in the literature or from our FEROS spectra.  The maximum
space velocity difference with respect to HIP\,65426 amounts to 12.9
km.s$^{-1}$.  Two stars, namely HIP\,64044 and TYC\,8653-1060-1, have
kinematic values very close to those of HIP\,65426 (space velocity
difference of 2.1 and 0.9 km.s$^{-1}$, respectively). However, being
at a projected separation larger than 2.9\,\degr\, they do not form a
bound system with HIP\,65426.  Fig.~\ref{f:isoc} shows the absolute
magnitude in $V$-band versus effective temperature diagram for
HIP\,65426 and stars within 5\,\degr\, with similar distance and
kinematics and comparison with theoretical models.  Effective
temperatures and reddening have been derived from spectral types using
the calibration of \cite{pecaut2013}.  The result of this comparison
shows that the typical age of LCC stars in the surroundings of
HIP\,65426 is of the order of 12-16 Myr.  This is fully consistent
with the Sco-Cen age map by \cite{pecaut2016}, that yields an age of
14\,Myr at the position of HIP\,65426 using a complementary approach
(much larger number of stars extending to fainter magnitude but
relying on kinematic parallaxes, while we used the Gaia trigonometric
values).  The nominal color-magnitude diagram age of HIP\,65426 is
younger (9-10 Myr). This is most likely due to the alteration induced
by its very fast rotational velocity \citep{david2015}.  Therefore, an
age of 14$\pm$4 Myr is adopted for this star.

\section{Stellar rotation }
\label{s:rot}

We derived radial and rotational velocity of HIP\,65426 from HARPS
spectra acquired during the nights of January 16, 17, and 18,
2017. The spectra have a resolution of 115000 and cover the spectral
range from 380 to 690\,nm. We obtained sequences of two, seven, and five
exposures of 5\,min each during the three nights. The spectra were
reduced using the HARPS pipeline that provides unidimensional spectra
sampled in uniform wavelength steps. We summed the spectra for each
night to provide higher S/N dataset minimizing the short-term
variations due to pulsations.

To test the methodology, we also applied the same procedure to
a number of additional A2V stars with archived HARPS spectra. The
results make use of a standard cross-correlation function method that
exploits a binary mask. We considered two sets of lines: (i) Six
strong lines (Ca II K and five H lines from H$\beta$ to H9, excluding
H$\epsilon$); we used the rotationally broadened core of the lines and
(ii) 35 atomic lines. The average radial velocity found ($5.2\pm1.3$
km.s$^{-1}$) is close to the literature value ($3.1\pm1.2$
km.s$^{-1}$). There is a small difference between the radial
velocities in the different dates (r.m.s. $\sim1.3$ km.s$^{-1}$).
However, we think that this result neither supports nor contradicts the
hypothesis that HIP\,65426 is a binary. We derived a very high
projected rotational velocity of $v \sin i
=299\pm9$\,km.s$^{-1}$. This value is at the upper limit of the
distribution of rotational velocities for A2V stars.  The extracted
values of RV and $v \sin i$ are consistent with literature values
within the error bars.

\begin{figure}[t]
\includegraphics[width=\columnwidth]{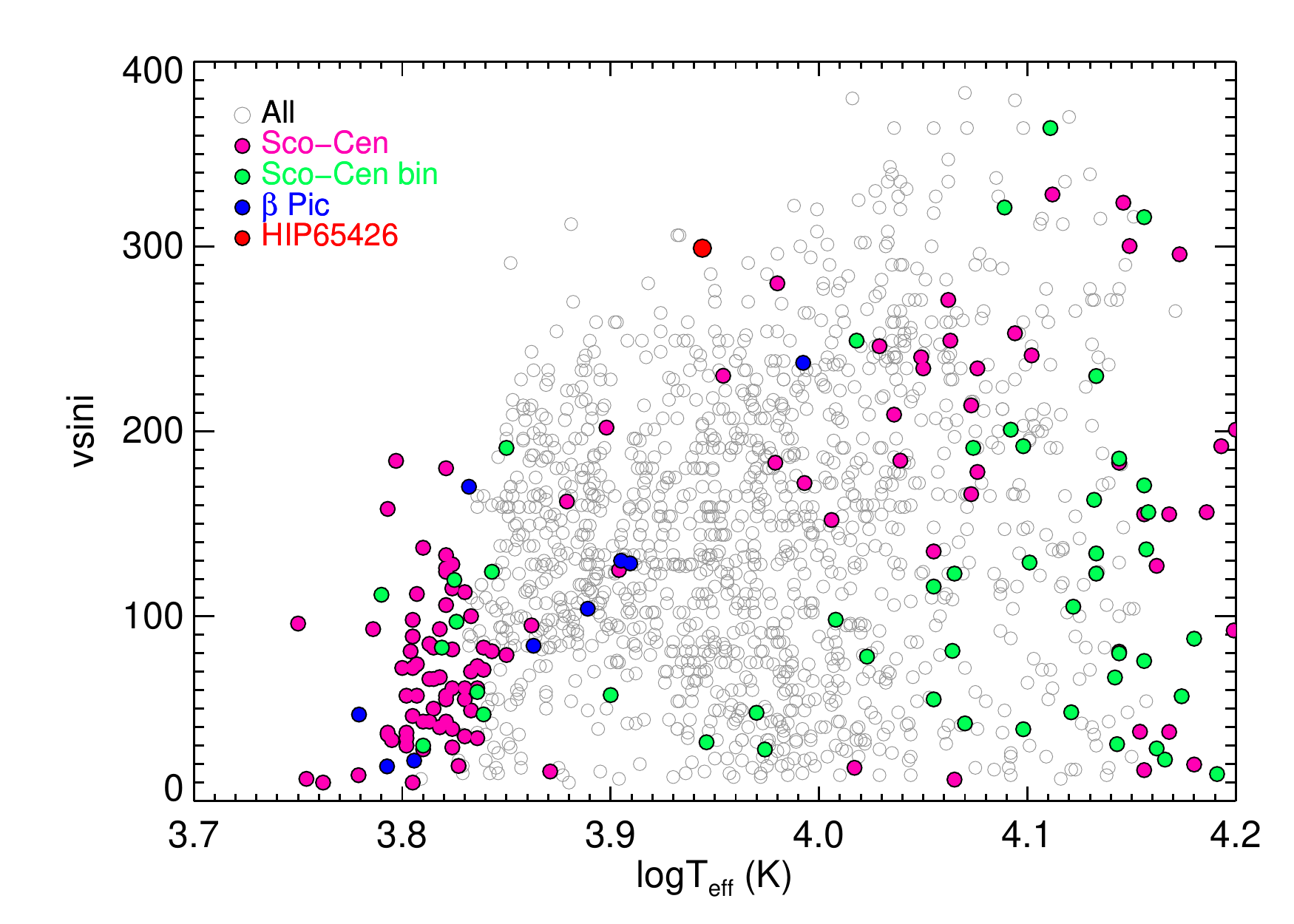}

\caption{Run of the $v \sin i$ values as a function of the effective
  temperature for “normal” stars in the catalog by Zorec et al.
  (2012: open gray circles). The \textit{red filled circle} is
  HIP\,65426. Superimposed are single (\textit{pink circles}) and binary (\textit{green
  circles}) stars in the Sco-Cen association. Stars in the $\beta$ Pic
  group are also shown (\textit{blue filled circles}).}

\label{fig:rot}
\end{figure}

If we examine the catalog by Zorec \& Royer (2012), there is no A2V
star with $v \sin i > 280$\,km.s$^{-1}$ among 119 entries. If we
extend the sample by one spectral subclass (A1V-A3V), there is only
one A1V star with a $v \sin i > 280$\,km.s$^{-1}$ (a close binary)
over 368 entries. HIP\,65426 is therefore an exceptionally fast rotator.
Fig.\,\ref{fig:rot} compares the $v \sin i$ value for HIP\,65426
with values obtained for “normal” stars (i.e., bonafide single stars)
in the Zorec et al. (2012) catalog. For comparison, we also plotted
single and binary stars in the Sco-Cen association and stars in the
$\beta$ Pic group that have similar ages. In these last cases, we also
considered $v \sin i$ values from Glebocki et al. (2005) and Chen
(2011) catalogs, after correcting them on the same scale as that of
the Zorec et al. catalog. The effective temperatures for these stars
were obtained from the B-V colours, after calibrating with those from
the Zorec et al. (2012) catalog. The peculiar nature of HIP\,65426 is quite
obvious from this figure. Also, note that the rotational velocities
for single late-B and early-A stars in the Sco-Cen association and in
the $\beta$ Pic group match well typical values for the “normal” stars
in the Zorec et al. (2012) catalog (while binaries typically rotate
slower, as also found for field stars). This suggests that stars with
masses similar or greater than HIP\,65426 have already reached the zero
age main sequence in these associations. HIP\,65426 is exceptional
with respect to both the Sco-Cen association and the field star
population.

\section{Binarity}
\label{s:bin}

HIP\,65426 is listed as a close visual binary (separation
0.15-0.3\,arcsec, $\Delta\sim0.1$~mag) in the Washington Double Star
Catalog \citep{wds}.  We retrieved the individual measurements (kindly
provided by Dr. B. Mason), that consist in seven entries between 1926
to 1933 followed by a series of non-detections.  There are no
indications of close companions in the SPHERE images, including the
non-coronagraphic sequence used for photometric calibrations, allowing
us to rule out the presence of equal-luminosity companions down to a
projected separation of 40 mas.  Furthermore, the HARPS spectra do not
show indications of multiple components.  Finally, an equal-luminosity
binary would imply an unphysical position on color-magnitude diagram below main
sequence. We thus consider it plausible that the previously claimed detection is
spurious and consider that we do not find any sign of binarity for HIP\,65426.

\section{Upper limits on the dust mass}
\label{s:excess}

\begin{figure}[ht]
\includegraphics[width=\columnwidth]{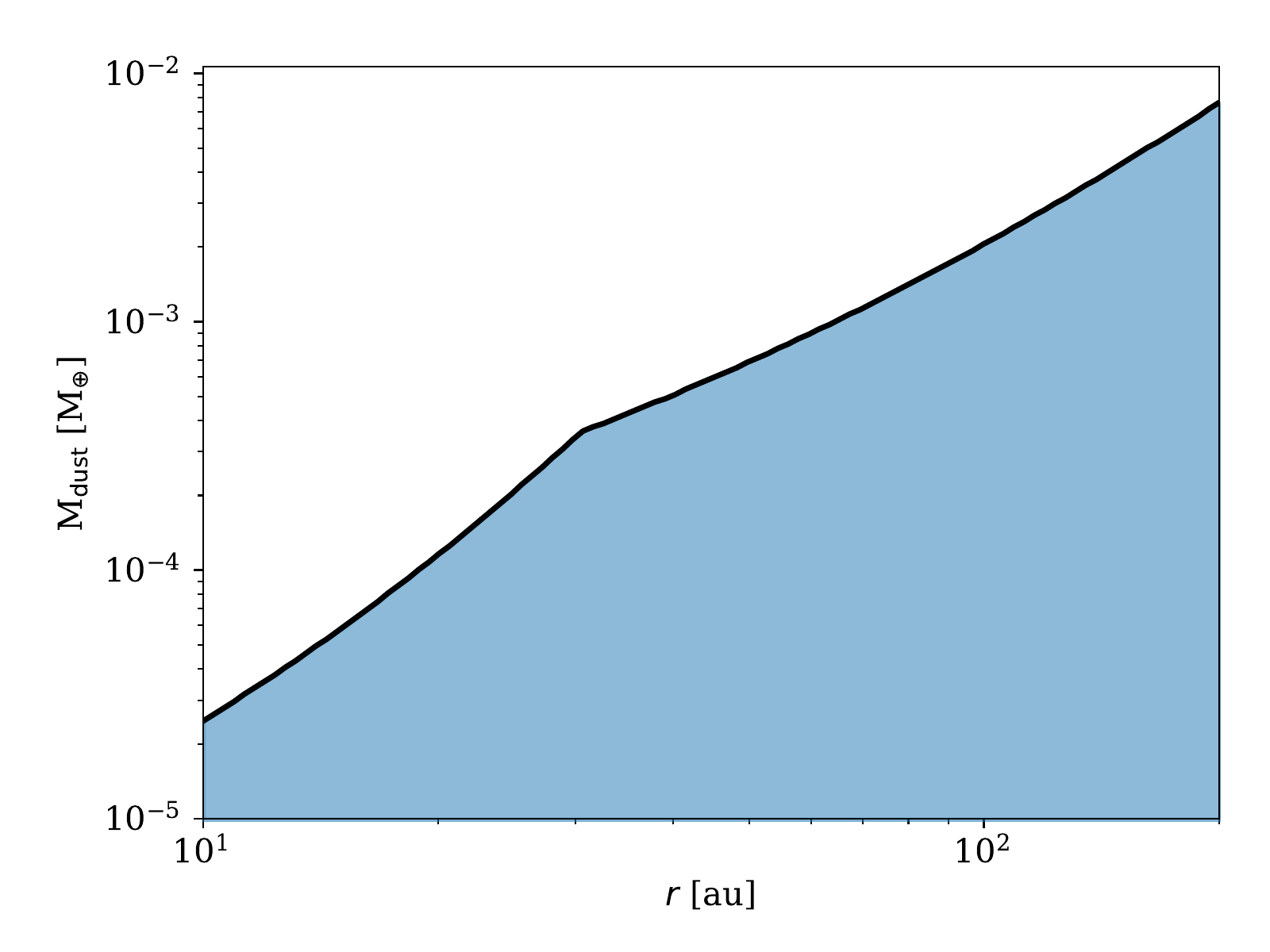}
\caption{The blue-shaded area shows the region in the
  $r$-$M_{\mathrm{dust}}$ plane, in which a debris disk could be
  present and remain compatible with the mid- and far-IR
  observations.}
\label{fig:mdust}
\end{figure}

\citet{chen2012} reported a non-detection of any mid-IR excess around
HIP\,65426. They reported a \textit{Spitzer}/MIPS upper limit at
70\,$\mu$m of $11.4$\,mJy. Given that the star is young, and that this
upper limit does not reach the photospheric flux ($\sim1.5$\,mJy), we
try to estimate the upper limit for the dust mass around HIP\,65426.
We gather the optical to mid-IR photometry of the star using
VOSA\footnote{http://svo2.cab.inta-csic.es/theory/vosa/}
\citep{Bayo2008}. For the given stellar luminosity and mass
($17.3\,L_{\odot}$ and $1.95\,M_{\odot}$, respectively) we estimate
the size of dust grains that would still be on bound orbits around the
star. We use optical constant of astro-silicates (\citep{Draine2003}),
and we compute the radiation pressure to gravitational forces $\beta$
ratio as in \citet{Burns1979}. We find that for this composition,
grains larger than $s_{\mathrm{blow}} \sim 3.1$\,$\mu$m should remain
on bound orbits around the star. To estimate the possible
configurations for a debris disk to remain compatible with the mid-
and far-IR observations, we compute a series of disk models (similar
to \citep{Olofsson2016}). We consider a grain size distribution of the
form d$n(s) \propto s^{-3.5}$d$s$, between $s_{\mathrm{min}} =
s_{\mathrm{blow}}$ and $s_{\mathrm{max}} = 1$\,mm. We sample $100$
$r_{\mathrm{i}}$ values for the radial distance between $10$ and
$200$\,au. For each $r_{\mathrm{i}}$, we consider a disk model between
$0.9 \times r_{\mathrm{i}} \leq r_{\mathrm{i}} \leq 1.1 \times
r_{\mathrm{i}}$. We then slowly increase the mass of the disk until
the thermal emission (plus the stellar contribution) is larger than
either the \textit{WISE}/W4 $22$\,$\mu$m point or the
\textit{Spitzer}/MIPS 70\,$\mu$m point. With this exercise we can
therefore delimit a region in the $r$-$M_{\mathrm{dust}}$ plane where
debris disks could exist and remain undetected with the current
observations (see Fig.\,\ref{fig:mdust}). Overall, with our
assumptions on the radial extent of the debris disk, we find that it
would have to be less massive that $10^{-3}$\,M$_{\oplus}$ at about
$100$\,au from the star.

\section{Observing Log}
\label{s:obslog}

\begin{table*}[ht]
\caption{Obs Log of VLT/SPHERE observations}             
\label{tab:obslog}
\centering
\begin{tabular}{lllllllllll}     
\hline\hline\noalign{\smallskip}
UT Date    &    Instr. & Filter   &  DIT$^a$  & $\rm{N}_{\rm{exp}}^a$ &   $\Delta\pi^a$     &  $\omega^a$    &   Strehl$^a$ & Airm.  & True North correction & Platescale \\ 
           &         &               &  (s)              &                       & ($\degr$)             & ($"$)   &  @1.6\,$\mu$m   &  &  ($\degr$) &   (mas/pixel)    \\
\noalign{\smallskip}\hline\noalign{\smallskip} 
30-05-2016 &    IFS & YJ     & 64 & 60  & \multirow{2}{*}{34.2} & \multirow{2}{*}{0.62 }  & \multirow{2}{*}{0.60 } & \multirow{2}{*}{1.12}   &    $-102.20\pm0.12$ &   $7.46\pm0.02$\\
30-05-2016 &    IRDIS & H2/H3 & 64 & 60  &                       &                         &                        &                         & $-1.72\pm0.06$    & $12.255/12.251^b\pm0.009$   \\
\noalign{\smallskip}
26-06-2016 &    IFS & YJ     & 64 & 75  & \multirow{2}{*}{42.2} & \multirow{2}{*}{0.53 }  & \multirow{2}{*}{0.66 } & \multirow{2}{*}{1.13}   &     $-102.25\pm0.12$  &   $7.46\pm0.02$\\
26-06-2016 &    IRDIS & H2/H3 & 64 & 75  &                       &                         &                        &                         & $-1.77\pm0.05$    & $12.255/12.251^b\pm0.009$   \\
\noalign{\smallskip}
07-02-2017 &    IFS & YJ     & 64 & 80  & \multirow{2}{*}{44.2} & \multirow{2}{*}{0.38 }  & \multirow{2}{*}{0.84 } & \multirow{2}{*}{1.13}    &     $-102.19\pm0.12$  &   $7.46\pm0.02$\\
07-02-2017 &    IRDIS & H2/H3 & 64 & 80  &                       &                         &                        &                         & $-1.71\pm0.05$    & $12.255/12.251^b\pm0.009$   \\
\noalign{\smallskip}
09-02-2017 &    IFS & YJH     & 64 & 80  & \multirow{2}{*}{49.1} & \multirow{2}{*}{0.45}  & \multirow{2}{*}{0.76 } & \multirow{2}{*}{1.14}   &    $-102.19\pm0.12$  &   $7.46\pm0.02$\\
09-02-2017 &    IRDIS & K1/K2  & 64 & 80  &                       &                         &                        &                         & $-1.71\pm0.05$    & $12.267/12.263^b\pm0.009$   \\
\noalign{\smallskip}
\hline\noalign{\smallskip}
\end{tabular}
\begin{list}{}{}
\item \scriptsize{(a) DIT refers to the integration time, $\rm{N}_{\rm{exp}}$ to the
  total number of frames in the final mastercube, $\Delta\pi$ to the parallactic angle range
  during the sequence. The seeing ($\omega$) and the Strehl ratio
  conditions were calculated by the SPHERE extreme-adaptive optics
  system. (b) The two values of platescale correpond to the platescale
  estimated for both dual-band filters indicated in the
  \textit{Filter} column.}
\end{list}
\end{table*}

\section{Astrometric and photometric detailed results}
\label{s:astrophot}

\begin{table}[h]
\caption{IRDIS relative astrometric measurements of HIP\,65426\,b to
  HIP\,65426 and IRDIS and IFS relative photometric contrast and
  absolute magnitudes for HIP\,65426\,b. The composite $J_{IFS}$-band is estimated between 1.20 and 1.32$~\mu$m.}
\label{tab:astro1}
\centering
\begin{tabular}{cccc}     
\hline\hline\noalign{\smallskip}
UT Date      & Filter          &   Separation (mas)    & P.A. (deg)             \\ \cline{1-2} \cline{3-4}
\noalign{\smallskip}
30/05/2016   & $H_2$       &   $830.4\pm4.9$ & $150.28\pm0.22$  \\
26/06/2016   &  $H_2$      &   $830.1\pm3.2$ & $150.14\pm0.17$  \\
07/02/2017   & $H_2$       &   $827.6\pm1.5$ & $150.11\pm0.15$  \\
09/02/2017   & $K_1$       &   $828.8\pm1.5$ & $150.05\pm0.16$  \\
\noalign{\smallskip}
\multicolumn{2}{c}{}              &   $\Delta$  (mag)     & $M_{abs}$  (mag)       \\ \cline{3-4}
\noalign{\smallskip}
07/02/2017 & $J_{2MASS}$           &   $12.67\pm0.40$  &  $14.26\pm0.42$                \\ 
           & $J_{IFS}$           &   $12.18\pm0.08$  &  $13.78\pm0.11$                              \\ 
07/02/2017 & $H_2$                &   $11.14\pm0.05$  &  $12.76\pm0.11$    \\ 
07/02/2017 & $H_3$                &   $10.78\pm0.06$  &  $12.39\pm0.12$    \\ 
09/02/2017& $K_1$                 &   $10.01\pm0.31$  &  $11.55\pm0.33$                 \\ 
09/02/2017& $K_2$                 &   $9.69\pm0.31$   &  $11.22\pm0.33$                 \\ 
\noalign{\smallskip}\hline\noalign{\smallskip} 
\end{tabular}
\end{table}

\begin{table}[h]
\caption{IRDIS $H_2$ relative astrometric and photometric measurements of June 26th, 2016 for the additional
  companion candidates in the field. }  
\label{tab:astro2}
\centering
\begin{tabular}{cccc}     
\hline\hline\noalign{\smallskip}
Cand.      &   Separation (mas)    & P.A. (deg) &   $\Delta$ (mag)           \\ 
\noalign{\smallskip}\hline\noalign{\smallskip} 
cc-1           &  $2494.7\pm12.2$          & $117.42\pm0.28$   &  $15.6\pm0.2$ \\
cc-2           &  $3973.1\pm14.4$          & $307.91\pm0.23$   &  $16.2\pm0.5$\\
cc-3           &  $6752.6\pm11.3$          & $304.28\pm0.15$   &  $14.1\pm0.2$\\
\noalign{\smallskip}
\hline
\end{tabular}
 \end{table}

\begin{figure}[ht]
\hspace{0.2cm}
\includegraphics[width=\columnwidth]{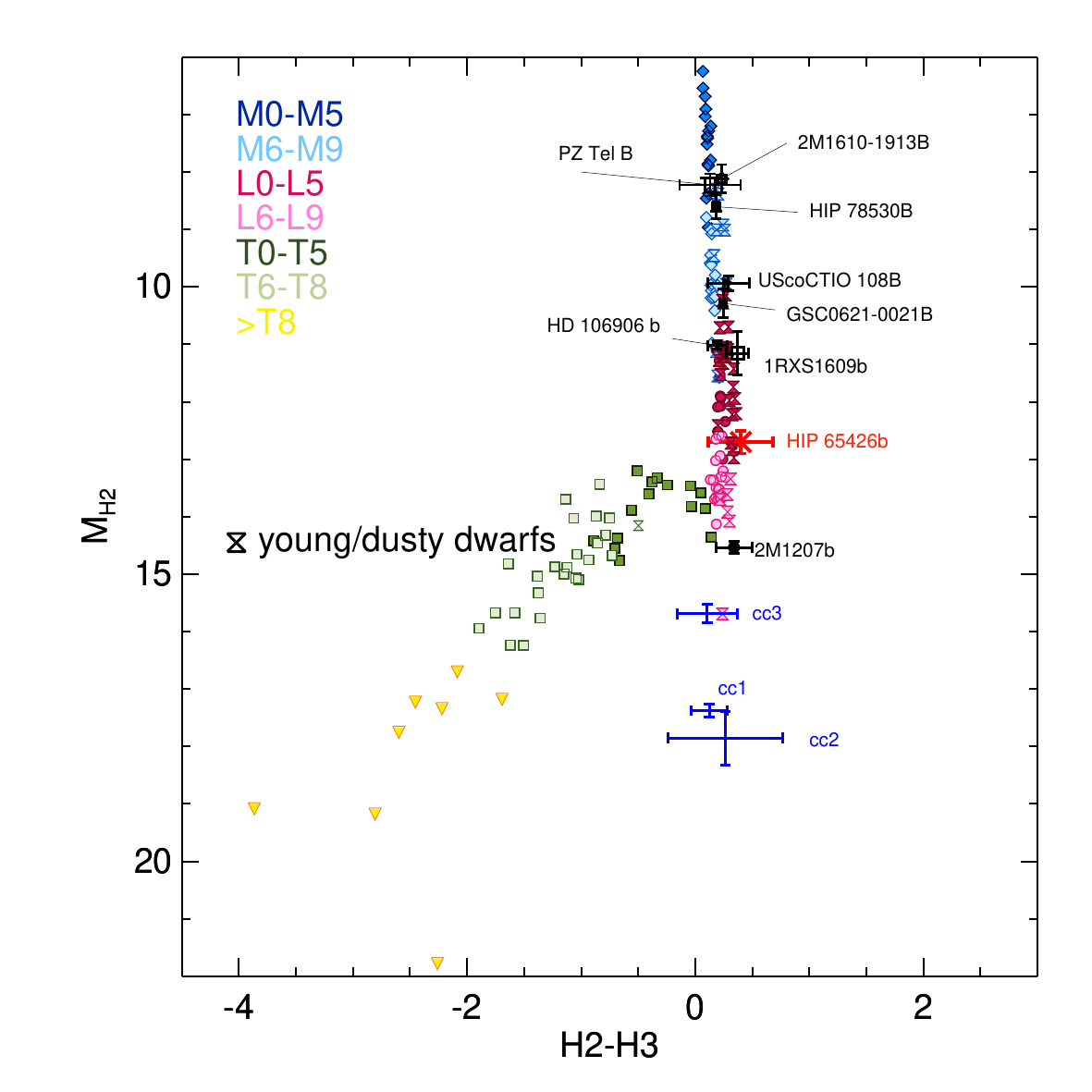}
\begin{centering}
\caption{Color-magnitude diagram considering the SPHERE/IRDIS $H_2$ and $H_3$ photometry. HIP\,65426 is indicated with error bars in \textit{red} and the other companion candidates are shown in \textit{blue}.}
\label{fig:cmd}
\end{centering}
\end{figure}

\end{document}